# Hydration Phase diagram for sodium cobalt oxide $Na_{0.3}CoO_2 \bullet yH_2O$

M.L. Foo[1], T. Klimczuk[1,3] and R.J. Cava[1,2]

[1]Department of Chemistry, Princeton University, [2]Princeton Institute for Research in Science and Technology of Materials, Princeton, NJ 08544, USA
[3]Faculty of Applied Physics and Mathematics, Gdansk University of Technology, Narutowicza 11/12, 80-952 Gdansk, Poland

The hydration phase diagram for sodium cobalt oxyhydrate, $Na_{0.3}CoO_2 \bullet yH_2O$ (y=0, 0.6, 1.3), was determined as a function of relative humidity at 298 K It is found that greater than 75 % relative humidity is needed for complete hydration of anhydrous $Na_{0.3}CoO_2$ to the superconducting phase $Na_{0.3}CoO_2 \bullet 1.3H_2O$. Dehydration studies show that a minimum of 43 % relative humidity is needed to maintain the stability of the fully hydrated superconducting phase. The intermediate hydrate, $Na_{0.3}CoO_2 \bullet 0.6H_2O$, is stable between 10 % and 50 % relative humidity on hydration, and 35 % to 0 % relative humidity on dehydration.

## Introduction

The discovery of superconductivity in $Na_{0.3}CoO_2 \bullet 1.3H_2O$ [1] with a critical temperature of 4.5 K is significant as its superconductivity may be derived from the special properties of a triangle-based magnetic lattice. Structurally, $Na_{0.3}CoO_2 \bullet 1.3H_2O$ consists of edge sharing $CoO_6$ octahedra in a triangular planar geometry, with interleaving planes of $H_2O$ molecules. The absence of superconductivity for both the lower hydrate, $Na_{0.3}CoO_2 \bullet 0.6H_2O$ and its parent anhydrous compound $Na_{0.3}CoO_2$ is intriguing. This demonstrates that superconductivity in cobaltates is strongly 2-D in nature because the correct inter-layer spacing of the $CoO_2$ layers must be present, not merely the appropriate electronic doping of Co.

When exposed to saturated water vapor or liquid water, anhydrous $Na_{0.3}CoO_2$ intercalates water between $CoO_2$ layers to form the superconducting bilayer hydrate $Na_{0.3}CoO_2 \bullet 1.3H_2O$. This induces expansion of the c-axis from c=11.2 Å to c=19.6 Å. The monolayer hydrate $Na_{0.3}CoO_2 \bullet 0.6H_2O$ can be formed by heating the bilayer hydrate to 35°C [2] or by exposing the superconductor to flowing dry $N_2$ at ambient temperature [3]. The monolayer hydrate has a c-axis of c = 13.8 Å, suggesting that the water molecules are in the same plane as the sodium since the Van der Waals diameter of a water molecule is ~2.8 Å . This has been confirmed by structural refinements of X-ray synchrotron data [3]. The three crystal structures of $Na_{0.3}CoO_2 \bullet yH_2O$, for y=0, 0.6 and 1.3, are shown schematically in Figure 1.

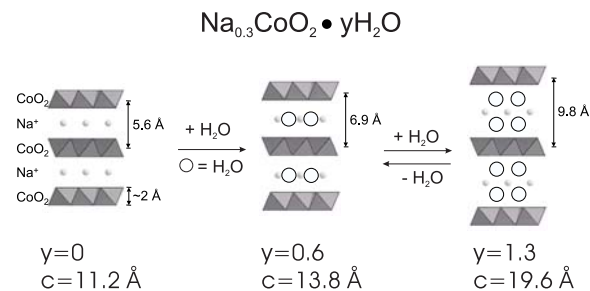

Figure 1: Schematic structures of sodium cobalt oxyhydrate $Na_{0.3}CoO_2 \bullet yH_2O$ (y=0, 0.6, 1.3) with emphasis on the $CoO_6$ octahedra. The interconversions of the hydrates at different relative humidities at room temperature are shown

The intercalation of water into layered inorganic compounds is not unique to the cobaltates. Hydration behavior of montmorillinite and vermiculite clays has been studied extensively [4]. For clays containing alkali ions such as $Li^+$ and $Na^+$, it has been observed that mono-, bi-, and even multi-layered hydrates can be formed [5]. The layered ternary sulfides such as $A_xMS_2$ where A= alkali metal; M=Ti, Nb, and Ta [6], are very similar to sodium cobalt oxide in terms of structure and hydration behavior. A careful gravimetric study on the hydration/dehydration behavior of $Na_{0.3}TaS_2 \bullet yH_2O$ has been performed by Johnston [7]. Unlike the cobalt oxides, anhydrous $Na_{0.3}TaS_2$ and its hydrates are superconducting with the same $T_c$ of 5K.

The preparation of the superconducting bilayer sodium cobalt oxyhydrate was originally reported to involve washing anhydrous $Na_{0.3}CoO_2$ with water and leaving the hydrate to dry at 'ambient' conditions [1]. In our experiments, we have observed that the formation of varying amounts of the monolayer hydrate occurs during drying. This implies that at room temperature (~298 K) the bilayer hydrate is extremely sensitive to dehydration that results from changes in relative humidity (RH). In this paper, we report the results of a systematic study of the hydration/dehydration behavior of $Na_{0.3}CoO_2 \bullet yH_2O$ as a function of RH at room temperature. The saturated salt solution method [8] is used to control RH due to its simplicity. The understanding of the hydration/dehydration properties of the sodium cobalt oxyhydrate superconductor developed in this work indicates guidelines for the handling and storage of samples to help ensure optimal properties.

**Experimental**

$Na_{0.7}CoO_2$ was synthesized by solid-state reaction of $Na_2CO_3$ (Alfa-Aesar, 99.5 %) and $Co_3O_4$ (Alfa-Aesar, 99.7 %) at 1073 K in flowing oxygen. Anhydrous $Na_{0.3}CoO_2$ was synthesized by reaction of $Na_{0.7}CoO_2$ with 40 X bromine in dry acetonitrile. (1 X is the theoretical amount of bromine needed to deintercalate all the sodium in $Na_{0.7}CoO_2$). The reaction was magnetically stirred for 5 days and washed with dry acetonitrile and dried under flowing argon gas.

For the hydration experiments, approximately 50 mg of anhydrous $Na_{0.3}CoO_2$ were placed in a 10 ml sealed chamber with the appropriate RH controlled by 3 ml of saturated salt solution (Table 1) at room temperature (298 K). Temperature fluctuations (± 2 °C) do not greatly affect the RH maintained by the saturated salt [8]. Dehydration experiments were performed using 50 mg of $Na_{0.3}CoO_2 \bullet 1.3H_2O$ employing the same setup as the hydration experiments. $Na_{0.3}CoO_2 \bullet 1.3H_2O$ was synthesized by washing $Na_{0.3}CoO_2$ in water, followed by drying in ambient and storage in a sealed chamber of 100 % relative humidity environment overnight. All samples were equilibrated for 2 weeks. Powder diffraction was performed using Cu $K_\alpha$ radiation operating in the continuous scan mode from 6-45° (2-theta) which allows rapid scanning (~5 min) to minimize sample degradation on exposure to ambient humidity.

| Relative Humidity (%) | Salt | Manufacturer / Purity |
|---|---|---|
| 0 | $P_2O_5$ | Fisher /99.8 % |
| 20 | $CaCl_2$ | EM /99.5 % |
| 33 | $MgCl_2 \bullet 6H_2O$ | Aldrich /99 % |
| 43 | $K_2CO_3$ | Alfa-Aesar /99 % |
| 52 | $Mg(NO_3)_2 \bullet 6H_2O$ | Aldrich /99 % |
| 66 | $NaNO_2$ | Aldrich /97 % |
| 75 | NaCl | Alfa-Aesar /99 % |
| 88 | $BaCl_2 \bullet 2H_2O$ | Alfa-Aesar /99 % |
| 93 | $NH_4H_2PO_4$ | Fisher /99% |
| 100 | ------ | ------ |

Table 1: Saturated salts used to maintain con trolled relative humidity [8, 14] at 298 K. For 100% relative humidity, no salt was used.

**Results and Discussion**

The X-ray patterns for hydration of anhydrous $Na_{0.3}CoO_2$ at different RH are shown in Figure 2. At ~0 % RH, $Na_{0.3}CoO_2$ does not intercalate water, however at 11 % RH, hydration to the monolayer hydrate $Na_{0.3}CoO_2 \bullet 0.6H_2O$ occurs. According to the phase rule, a two phase region must exist between these two RHs. Further experiments between these two RHs must be carried out to determine the exact location of the phase boundary. The monolayer hydrate persists as a single phase until a RH of 52 %. The (00$l$) peaks are asymmetrically broadened towards smaller d-spacing at this RH, indicating the presence of monohydrates with lower water content [2]. Between 52 % and 75 % RH, a distinct two-phase region exists with the lower hydrate in equilibrium with the bilayer hydrate $Na_{0.3}CoO_2 \bullet 1.3H_2O$. Above 75 % RH, the bilayer hydrate exists as a single phase, with the narrow peak width of the (00$l$) peaks indicative of a line composition. There is no further expansion of d-spacing for the (00$l$) peaks for higher RHs, indicating that a previously reported 'over-hydrated phase' $Na_{0.3}CoO_2 \bullet 1.8H_2O$ [9] is not observed in these experiments.

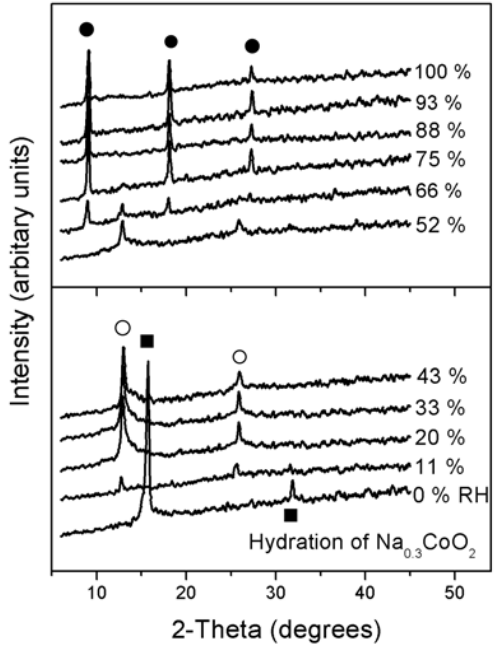
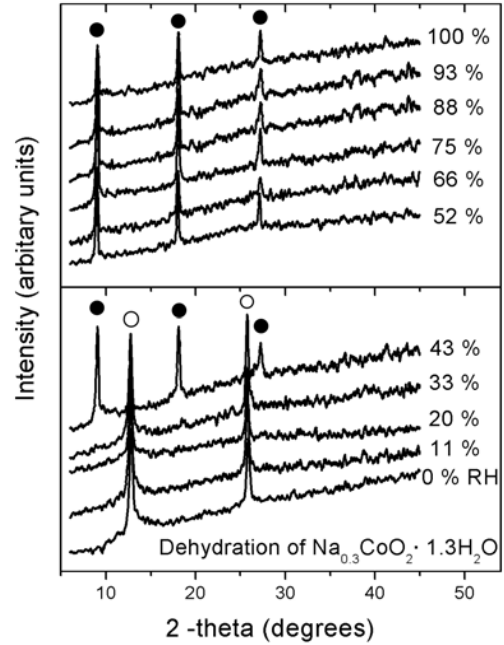

Figure 2: X-ray powder diffraction of anhydrous $Na_{0.3}CoO_2$ hydrated at different relative humidities for 2 weeks. (●-$Na_{0.3}CoO_2\bullet1.3H_2O$, ○- $Na_{0.3}CoO_2\bullet0.6H_2O$, ■- $Na_{0.3}CoO_2$). Peaks are indexed as (002), (004) and (006) from space group $P6_3/mmc$ (194).

Figure 3: X-ray powder diffraction of bilayer hydrate $Na_{0.3}CoO_2\bullet1.3H_2O$ dehydrated at different relative humidities for 2 weeks. (●-$Na_{0.3}CoO_2\bullet1.3H_2O$, ○- $Na_{0.3}CoO_2\bullet 0.6H_2O$) Peaks are indexed as (002), (004) and (006) from space group $P6_3/mmc$ (194).

The results of the dehydration experiments on $Na_{0.3}CoO_2\bullet1.3H_2O$ are presented in Figure 3. The bilayer hydrate is first found to be stable at 43 % RH, but at 33 % RH only the monolayer hydrate is observed. Thus, in comparison to the hydration experiments, no two phase region consisting of the monolayer and bilayer hydrate was directly observed by X-ray diffraction. Therefore it is concluded that this two-phase region should exist between the 43 % and 33 % RH. This two phase region is noticeably smaller than that observed for hydration of $Na_{0.3}CoO_2$. These results demonstrate that the bilayer hydrate, once prepared, can be stored at 43 % RH without dehydration to its lower hydrate. The lower hydrate $Na_{0.3}CoO_2\bullet0.6H_2O$ is observed to be stable even at 0 % RH on dehydration. Anhydrous $Na_{0.3}CoO_2$ was not obtained. Similar to the results in the hydration experiments on $Na_{0.3}CoO_2$, there are no obvious shifts in the (00$l$) peaks for either the lower or the bilayer hydrates equilibrated at different RHs, suggesting that there is no variation in water content for the hydrates under different conditions. Our structural characterization is not sufficiently precise to determine whether intergrowth of y=0, 0.6, and 1.3 layers occurs in substantial amounts at intermediate RHs, such as is seen in the hydrated $TaS_2$ materials [10].

The phase diagram for hydration/dehydration of anhydrous $Na_{0.3}CoO_2$/ $Na_{0.3}CoO_2\bullet1.3H_2O$ is shown in Figure 4. Hysteresis is observed since the RH for obtaining single phase bilayer hydrate through hydration (88 % RH) is markedly higher than its stability limit on dehydration (43 % RH). We attribute this to an energy cost for dehydration [11] derived from hydrogen bonding between the apical oxygens in the $CoO_2$ layers and hydrogen in the intercalated water molecules. In $Na_{0.3}TaS_2\bullet yH_2O$ this hydrogen bonding interaction is absent, which may explain why no significant hysteresis was observed [7].

From the hydration studies, we conclude that anhydrous $Na_{0.3}CoO_2$ should be kept in a dessicator filled with $P_2O_5$ or in an argon filled glovebox to prevent hydration. From the dehydration studies, during the 'drying' stage of synthesis, $Na_{0.3}CoO_2 \bullet 1.3H_2O$ is unstable, and dehydrates to the intermediate phase if the RH is less than 43 %. Hence overnight storage in a 100 % humidity chamber is recommended to produce the fully hydrated bilayer superconductor. Hydration of $Na_{0.3}CoO_2$ in a lower humidity than 100 % is not recommended. However, we note that prolonged storage of samples in 100 % RH environment causes condensation of water between the grains. The effect of this adsorbed and intergranular water must be considered when performing experiments on these samples. For optimum sample handling, prompt freezing of properly hydrated superconducting samples is suggested. We have observed that samples with good bulk superconductivity, when stored at liquid nitrogen temperature, have remained stable for at least six months with the same $T_c$ of 4.3 K. If necessary, the intergranular water (~15-20 % by mass) [12] can be eradicated by dynamic pumping of the frozen hydrated sample [13].

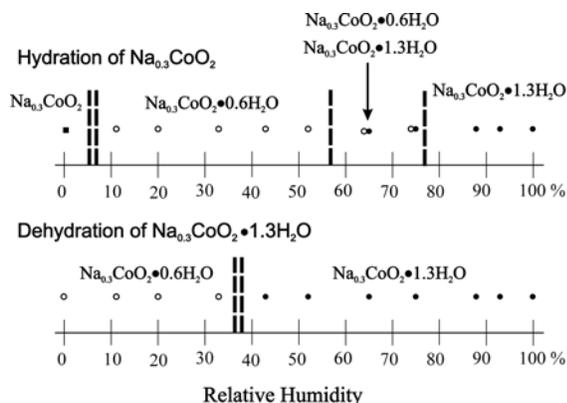

Figure 4: Hydration phase diagram of $Na_{0.3}CoO_2 \bullet yH_2O$. Vertical dashed lines represent phase boundaries.

**Acknowledgements**
This work was supported by the US Department of Energy, grant DE-FG02-98-ER 45706.